\documentclass{tlp}

 \usepackage{epic}
 \usepackage{eepic}
 \usepackage{epsfig}
 \usepackage{url}
 \usepackage{xspace}
 \usepackage{latexsym}
 \usepackage{amsmath}
 \usepackage{amssymb}

\newif\ifreporxt

\newcommand{\TODO}[1]{{\large\bf TODO: }#1\ensuremath{\Box}}
\newcommand{\nop}[1]{}

\newcommand{\Or}{\ensuremath{\mathtt{\,v\,}}\xspace}
\newcommand{\derives}{\ensuremath{\mathtt{\ :\!\!-}\ }}

\newcommand{\p}{\ensuremath{{\mathcal{P}}}}
\newcommand{\GP}{\ensuremath{Ground(\p)}}

\newcommand{\BP}{\ensuremath{B_{\p}}}
\newcommand{\UP}{\ensuremath{U_{\p}}}

\newcommand{\intelligent}{\ensuremath{\p}}

\newcommand{\R}{\ensuremath{r}}

\newcommand{\HR}{\ensuremath{H(\R)}}
\newcommand{\BR}{\ensuremath{B(\R)}}

\newcommand{\naf}{\ensuremath{\mathtt{not}}\xspace}

\newcommand{\dlv}{{\sc DLV}\xspace}

\newcommand{\tuple}[1]{\langle#1\rangle}

\newcommand{\q}{\ensuremath{{\cal Q}}}

\newcommand{\ground}[1]{\ensuremath{Ground(#1)}}

\newcommand{\bravecons}{\ensuremath{\models_b}}
\newcommand{\cautiouscons}{\ensuremath{\models_c}}

\newcommand{\head}[1]{\ensuremath{H(#1)}}
\newcommand{\body}[1]{\ensuremath{B(#1)}}
\newcommand{\posbody}[1]{\ensuremath{B^+(#1)}}
\newcommand{\negbody}[1]{\ensuremath{B^-(#1)}}

\newcommand{\atoms}[1]{\ensuremath{Atoms(#1)}}

\newenvironment{simpleprogram}[1][]
   {\vspace{-0.5ex}\begin{itemize}\item[]
      \tt
      \begin{tabbing}
      \code{#1}\ \= \kill
   }
   {\end{tabbing}\end{itemize}\vspace{-3ex}}

\newenvironment{simplealignedprogramstub}[1][]
   {\vspace{-0ex}
      \begin{tabbing}
      #1\kill
   }
   {\end{tabbing}
\vspace{-6ex}}

\newenvironment{sublabeledprogram}[1][]
   {\begin{array}{ll}\setlength{\arraycolsep}{0pt}}
   {\end{array}}

\newcommand{\code}[1]{\ensuremath{#1}}
\newenvironment{dlvcode}
  {\begin{displaymath}\begin{array}{l}}
  {\end{array}\end{displaymath}}
\newenvironment{dlvcode2}
  {\begin{displaymath}\begin{array}{ll}}
  {\end{array}\end{displaymath}}

\newtheorem{theorem}{Theorem}[section]
\newtheorem{example}[theorem]{Example}
\newtheorem{definition}[theorem]{Definition}
\newtheorem{proposition}[theorem]{Proposition}
\newtheorem{corollary}[theorem]{Corollary}
\newtheorem{lemma}[theorem]{Lemma}

\renewcommand{\P}{\mathcal{P}}
\newcommand{\Q}{\mathcal{Q}}
\newcommand{\SM}{\mathcal{SM}}

\newcommand{\qrelation}[3]{\ensuremath{{#1}_{#2}^{#3}}}

\newcommand{\qequiv}[2]{\ensuremath{\qrelation{\equiv}{#1}{#2}}}
\newcommand{\bqequiv}[1]{\ensuremath{\qequiv{#1}{b}}}
\newcommand{\cqequiv}[1]{\ensuremath{\qequiv{#1}{c}}}

\newcommand{\magicRules}{\ensuremath{\mathit{magicRules}}}
\newcommand{\modifiedRules}{$\mathit{modifiedRules}$}

\newcommand{\dmsqp}{\ensuremath{\DMS(\Q,\p)}}

\newcommand{\killed}[4]{\ensuremath{\mathtt{killed}^{#1}_{#3,#4}(#2)}}
\newcommand{\killedmpmp}{\ensuremath{\killed{M'}{M'}{\Q}{\p}}}
\newcommand{\killedmpnp}{\ensuremath{\killed{M'}{N'}{\Q}{\p}}}
\newcommand{\killedmn}{\ensuremath{\killed{M}{N}{\Q}{\p}}}

\newcommand{\variant}[3]{\ensuremath{\mathtt{variant}_{#1,#2}(#3)}}
\newcommand{\variantqpm}{\ensuremath{\variant{\Q}{\p}{M}}}
\newcommand{\variantqpi}{\ensuremath{\variant{\Q}{\p}{I}}}

\newcommand{\magica}{*}

\newcommand{\DMS}{\ensuremath{\mathtt{DMS}}}

\renewcommand{\t}{\bar t}
\newcommand{\s}{\bar s}

\def\<{\mbox{$\langle$}}
\def\>{\mbox{$\rangle$}}

\newcounter{myenumctr}

\newcommand{\EDB}{\ensuremath{E\!D\!B}\xspace}
\newcommand{\IDB}{\ensuremath{I\!D\!B}\xspace}
\newcommand{\Facts}{\ensuremath{Facts}}

\newcommand{\fg}{$\mathcal{FG}$\xspace}

\newcommand{\dquo}[1]{``#1"}

\newcommand{\implied}{\derives}

\newcommand{\hs}{\hspace{3mm}}

\newcommand{\ASPFN}{\ensuremath{\rm ASP^{\rm fs}}}
\newcommand{\ASPFNFR}{\ensuremath{\rm ASP^{\rm fs}_{\rm fr}}}

\newcommand{\blank}{\ensuremath{{\empty}_\sqcup}}

\begin{document}
\bibliographystyle{acmtrans}

\title[Disjunctive ASP with Functions: Decidable Queries and Effective Computation\ \ \ \ ]
    {Disjunctive ASP with Functions:\\ Decidable Queries and Effective Computation\thanks{This research has been partly supported by Regione Calabria and EU
under POR Calabria FESR 2007-2013 within the PIA
project of DLVSYSTEM s.r.l., and by MIUR under the PRIN project LoDeN. }
}

\author[M. Alviano, W. Faber and N. Leone]
{MARIO ALVIANO, WOLFGANG FABER and NICOLA LEONE\\
Department of Mathematics, University of Calabria \\
87036 Rende (CS), Italy\\
\email{\{alviano,faber,leone\}@mat.unical.it}
}

\date{}

\submitted{8 February 2010}
\revised{1 May 2010}
\accepted{16 May 2010}

\maketitle

\label{firstpage}

\begin{abstract}
Querying over disjunctive ASP with functions is a
highly undecidable task in general.
In this paper we focus on disjunctive logic programs with stratified
negation and functions under the stable model semantics (\ASPFN). We show that
query answering in this setting is decidable, if the query is
finitely recursive (\ASPFNFR).
Our proof yields also an effective method for query evaluation.  It is
done by extending the magic set technique to \ASPFNFR{}.  We show that
the magic-set rewritten program is query equivalent to the original
one (under both brave and cautious reasoning).
Moreover, we prove that the rewritten program is also finitely ground,
implying that it is decidable.  Importantly, finitely ground programs
are evaluable using existing ASP solvers, making the class of \ASPFNFR{} queries usable in practice.
\end{abstract}

\begin{keywords}
 answer set programming, decidability, magic sets, disjunctive logic programs
\end{keywords}

\section{Introduction}

Answer Set Programming (ASP), Logic Programming (LP) under the answer set
or stable model semantics, has established itself as a convenient and
effective method for declarative knowledge representation and
reasoning over the course of the last 20 years
\cite{bara-2002,gelf-lifs-91}. A major reason for the success of ASP
has been the availability of implemented and efficient systems, which
allowed for the paradigm to be usable in practice.

This work is about ASP with stratified negation
and functions under the stable model semantics (\ASPFN). Dealing with
the introduction of function symbols in the language of ASP has been
the topic of several works in the
literature~\cite{bona-02-iclp,bona-04,base-etal-2009-tplp,cali-etal-2009-lpnmr,syrj-2001,gebs-etal-2007-lpnmr,cali-etal-2008-iclp,lier-lifs-2009-iclp,simk-eite-2007-lpar,eite-simk-2009-ijcai,lin-wang-2008-KR,caba-2008-iclp}.
They have been motivated by overcoming the major limitation of ASP systems with respect to traditional LP systems, which is the possibility of representing only a finite set of individuals by means of constant symbols. Most of the approaches treat
function symbols in the traditional logic programming way, that is by
considering the Herbrand universe. A few other works treat function
symbols in a way which is closer to classical logic (see, e.g., \cite{caba-2008-iclp}). 
The fundamental problem with admitting function symbols in ASP is that
the common inference tasks become undecidable. The identification of
expressive decidable classes of ASP programs with functions is
therefore an important task, and has been addressed in several works
(see Section~\ref{sec:relatedwork}). 

Here, we follow the traditional logic programming approach, and study
the rich language of finitely recursive \ASPFN
(\ASPFNFR), showing that it is still decidable.
In fact, our work links two 
relevant
classes of ASP with functions: finitely recursive and
finitely ground programs. We extend a magic set method for programs
with disjunctions and stratified negation to deal with functions and
specialize it for finitely recursive queries. We show that the
transformed program is query equivalent to the original one and that
it belongs to the class of finitely ground programs. Finitely ground
programs have been shown to be decidable and therefore it follows that
\ASPFNFR{} queries are decidable, too.
Importantly, by \dlv-Complex~\cite{dlvcomplex-web}
there is a system which supports query answering on finitely ground
programs, so the magic set method serves also as a means for
effectively evaluating \ASPFNFR{} queries. We also show that
\ASPFNFR{} programs are maximally expressive, in the sense that each
computable function can be represented. In total, \ASPFNFR{} programs
and queries are an appealing formalism, since they are decidable, a
computational system exists, they provide a rich knowledge-modeling
language, including disjunction and stratified negation, and they can express any
computable function.

Summarizing, the main contributions of the paper are the following:
\begin{itemize}
\item[$\blacktriangleright$] 
\vspace{-0.2cm}We prove that \ASPFNFR{} queries are decidable
under both brave and cautious reasoning.
\item[$\blacktriangleright$] 
We show that the restrictions which guarantee the
decidability of \ASPFNFR{} queries do not limit
their expressiveness.
Indeed, we demonstrate that any computable function can be expressed
by an \ASPFNFR{} program.
\item[$\blacktriangleright$] 
We provide an effective implementation method for 
\ASPFNFR{} queries, making reasoning over \ASPFNFR{} programs 
feasible in practice. In particular,
\begin{itemize}
\item
\vspace{-0.2cm}We design a magic-set rewriting technique for \ASPFNFR{}
queries.
The technique is based on a particular {\em sideways information
passing strategy} (SIPS) which exploits the
structure of \ASPFNFR{} queries, and guarantees that
the rewritten program has a specific shape.
\item
We show that the magic-set rewritten program is query equivalent to the original
one (under both brave and cautious reasoning).
\item
We prove that the rewritten program is finitely ground,
implying that it is computable~\cite{cali-etal-2008-iclp}.
Importantly, finitely ground programs are evaluable using the existing ASP solver \dlv-Complex \cite{dlvcomplex-web}, making 
\ASPFNFR{} queries usable in practice.
\end{itemize}
\end{itemize}

\section{Preliminaries}\label{sec:preliminaries}

In this section, we recall the basics of ASP with function symbols, and
the decidable classes of finitely ground~\cite{cali-etal-2008-iclp}
and finitely recursive programs~\cite{base-etal-2009-tplp}.

\subsection{ASP Syntax and Semantics}

A \emph{term} is either a \emph{variable} or a \emph{functional
term}. 
A functional term is of the form $f\tt(t_1, \dots, t_k)$, where
$f$ is a function symbol ({\em functor}) of arity $k \ge 0$,
and $\tt t_1, \ldots, t_k$ are terms\footnote{We also use Prolog-like square-bracketed list notation
as in~\cite{cali-etal-2008-iclp}.}.
A functional term with arity 0 is a {\em constant}.
If $\tt p$ is a {\em predicate} of arity $k \geq 0$,
and $\tt t_1, \ldots, t_k$ are terms,
then $\tt p(t_1, \ldots, t_k)$ is an {\em atom}\footnote{%
We use the notation $\tt \t$ for a sequence
of terms, 
for referring to atoms as
$\tt p(\t)$.}. A {\em literal} is either an atom $\tt p(\t)$ (a positive literal),
or an atom preceded by the {\em negation as failure} symbol $\tt \naf~p(\t)$
(a negative literal).
A {\em rule} $\R$ is of the form
\begin{dlvcode}
\tt p_1(\t_1) \ \Or\ \cdots \ \Or\ p_n(\t_n) \derives 
    q_1(\s_1),\ \ldots,\ q_j(\s_j),\ \naf~q_{j+1}(\s_{j+1}),\ \ldots,\ \naf~q_m(\s_m).
\end{dlvcode}%
where $\tt p_1(\t_1),\ \ldots,\ p_n(\t_n),\ q_1(\s_1),\ \ldots,\ q_m(\s_m)$ 
are atoms and $n\geq 1,$  $m\geq j\geq 0$. The
disjunction $\tt p_1(\t_1) \ \Or\ \cdots \ \Or\ p_n(\t_n)$ is the {\em head} of~\R{}, 
while the conjunction 
$\tt q_1(\s_1),\ \ldots,\ q_j(\s_j),\ \naf~q_{j+1}(\s_{j+1}),\ \ldots,\ \naf~q_m(\s_m)$ 
is the {\em body} of~\R{}.
Moreover, $\HR$ denotes the set of head atoms, while $\BR$ denotes the set of body literals.
We also use $\posbody{\R}$ and $\negbody{\R}$ for denoting
the set of atoms appearing in positive and negative body literals, respectively,
and $\atoms{\R}$ for the set $\HR \cup \posbody{\R} \cup \negbody{\R}$.
A rule $\R$ is normal (or disjunction-free) if $|\HR| = 1$, 
positive (or negation-free) if $\negbody{\R}=\emptyset$,
a {\em fact} if both $\body{\R}=\emptyset$, 
$|\HR| = 1$ and no variable appears
in $\HR$.

A \emph{program}
$\P$ is a finite set of rules; if all the rules in it are positive (resp.\ normal), 
then $\P$ is a positive (resp.\ normal) program.
In addition, $\p$ is function-free if each functional term
appearing in $\p$ is a constant.
Stratified programs constitute another interesting class of 
programs. A predicate $\tt p$ appearing in the head of a rule $\R$
{\em depends} on each predicate $\tt q$ such that an atom $\tt q(\s)$ belongs to $\BR$; 
if $\tt q(\s)$ belongs to $\posbody{\R}$, $\tt p$ depends on $\tt q$ positively, otherwise negatively. 
A program is \emph{stratified} if there is no cycle of dependencies involving a
negative dependency.
In this paper we focus on the class of stratified programs.

Given a predicate $\tt p$, a {\em defining rule} for $\tt p$ is a rule
$\R$ such that some atom $\tt p(\t)$ belongs to $\head{\R}$. If all
defining rules of a predicate $\tt p$ are facts, then $\tt p$ is an
\EDB\ {\em predicate}; otherwise $\tt p$ is an \IDB\ {\em
predicate}\footnote{\EDB\ and \IDB stand for Extensional
Database and Intensional Database, respectively.}. 
Given a program $\p$,  
the set of rules having some IDB predicate in head
is denoted by $\IDB(\p)$, while $\EDB(\p)$ denotes the remaining rules, that is,
$\EDB(\p) = \P \setminus \IDB(\p)$.
In addition, the set of
all facts of $\p$ is denoted by $\Facts(\p)$.

The set of terms constructible by combining functors appearing in a program $\P$ is 
the \emph{universe} of $\P$ and is denoted by $\UP$,
while the set of ground atoms constructible from predicates in $\P$ with elements of $\UP$ is the \emph{base}
of $\P$, denoted by $\BP$. We call a term (atom, rule, or program) 
\emph{ground} if it does not contain any variable.
A ground atom $\tt p(\t)$ (resp.\ a ground rule $\R_g$) is
an instance of an atom $\tt p(\t')$ (resp.\ of a rule $\R$) if there is a 
substitution $\vartheta$ from the variables in $\tt p(\t')$ (resp.\ in $\R$) 
to $\UP$ such that ${\tt p(\t)} = {\tt p(\t')}\vartheta$ 
(resp.\ $\R_g = \R\vartheta$). 
Given a program $\p$, $\ground{\p}$ denotes the set of all the instances
of the rules in $\p$.

\nop{
Given an atom $\tt p(\t)$ and a set of ground atoms $A$,
by $A|_{\tt p(\t)}$ we denote the set of ground instances of $\tt p(\t)$ belonging to $A$.
For example, $\BP|_{\tt p(\t)}$ is the set of all the ground atoms obtained by applying to
$\tt p(\t)$ all the possible substitutions from the variables in $\tt p(\t)$ to $\UP$,
that is, the set of all the instances of $\tt p(\t)$. 
Abusing of notation, if $B$ is a set of atoms, by $A|_B$ we denote the union
of $A|_{\tt p(\t)}$, for each ${\tt p(\t)} \in B$.
}

An \emph{interpretation} $I$ for a program $\P$ is a subset of $\BP$. A positive ground 
literal $\tt p(\t)$ is true w.r.t.\ an
interpretation $I$ if ${\tt p(\t)}\in I$; otherwise, it is false. 
A negative ground literal $\tt \naf\ p(\t)$ is true w.r.t.\ $I$ 
if and only if $\tt p(\t)$ is false w.r.t.\ $I$. 
The body of a ground rule $\R_g$ is true w.r.t.\ $I$
if and only if all the body literals of $\R_g$ are true w.r.t.\ $I$, that is,
if and only if $\posbody{\R_g} \subseteq I$ and $\negbody{\R_g} \cap I = \emptyset$. 
An interpretation $I$ {\em satisfies} a ground rule $\R_g\in \GP$ if at least one atom
in $\head{\R_g}$ is true w.r.t.\ $I$ whenever the body of $\R_g$ is true w.r.t.\ $I$. An interpretation $I$ is a
\emph{model} of a program $\P$ if $I$ satisfies all the rules in $\GP$. 
\nop{
Since an interpretation is a set of atoms, if $I$ is an interpretation
for a program $\p$, and $\p'$ is another program,
then by $I|_{B_{\p'}}$ we denote the restriction of $I$ to the predicate
symbols in $\p'$.
}

Given an interpretation $I$ for a program $\P$, the reduct of $\P$ w.r.t.\ $I$,
denoted $\ground{\p}^{I}$, is obtained by deleting from $\GP$ all 
the rules $\R_g$ with $\negbody{\R_g} \cap I = \emptyset$,
and then by removing all the negative literals from the remaining rules.
The semantics of a program $\P$ is then given by the set $\SM(\P)$ of the stable models of
$\P$, where an interpretation $M$ is a stable model for $\P$ if and only if
$M$ is a subset-minimal model of $\ground{\P}^M$.

Given a program $\p$ and a query $\Q = {\tt g(\t)?}$ (a ground atom)%
\footnote{More complex queries can still be expressed using appropriate rules.
We assume that each functor appearing in $\q$ also appears in $\p$;
if this is not the case, then we can add to $\p$ a fact $\tt p(\t)$
(where $\tt p$ is a predicate that occurs neither in $\p$ nor $\Q$)
and $\tt \t$ are the arguments of $\q$.}, 
$\p$ {\em cautiously} (resp.\ {\em bravely}) entails $\Q$, denoted $\p \cautiouscons \Q$ (resp.\ $\p \bravecons \Q$) if and only if
${\tt g(\t)} \in M$ for all (resp.\ some) $M \in \SM(\P)$.
Two programs $\p$ and $\p'$ are \emph{cautious-equivalent} (resp.\ \emph{brave-equivalent}) 
w.r.t.\ a query $\Q$, 
denoted by $\P\cqequiv{\Q} \P'$ (resp.\ $\P\bqequiv{\Q} \P'$), 
whenever $\p \cautiouscons \Q$ iff $\p' \cautiouscons \Q$
(resp.\ $\p \bravecons \Q$ iff $\p' \bravecons \Q$).

\subsection{Finitely Ground Programs}\label{subsec:fg_programs}
The class of finitely ground (\fg)
programs~\cite{cali-etal-2008-iclp} constitutes a natural
formalization of programs which can be finitely evaluated bottom-up.
\nop{Informally, the definition of finitely ground program relies on
the so-called \dquo{intelligent instantiation}, obtained by
means of an operator which is iteratively applied on program's
submodules, producing sets of ground rules. In order to
properly split a given program $\p$ into modules, it is taken
in consideration the {\em dependency graph} and the {\em
component graph}. The first connects predicate names, 
while the latter connects strongly
connected components of the former. Each module corresponds to
a strongly connected component (SCC)\footnote{We recall here
that a strongly connected component of a directed graph is a
maximal subset $C$ of the vertices, such that each vertex in
$C$ is reachable from all other vertices in $C$.} of the
dependency graph. An ordering relation is then defined among
modules/components: a {\em component ordering} $\gamma$ for
$\p$ is a total ordering such that the intelligent
instantiation $\p^\gamma$, obtained iteratively by following the
sequence given by $\gamma$, has the same stable models of
$\GP$.


For the sake of clarity, we shortly recall here some key
concepts introduced in~\cite{cali-etal-2008-iclp}. 
For complete formal definitions,
more details and examples, we refer the reader to the
aforementioned paper. 
}
We recall the key concepts, and refer to~\cite{cali-etal-2008-iclp} for details and examples.

The dependency graph ${\cal G}(\p)$ of a program $\p$
is a directed graph having a node for each
IDB predicate of $\p$, and an edge $\tt q \rightarrow p$ if 
there is a rule $\R \in \p$ such that $\tt p$ occurs in $\HR$ and 
$\tt q$ occurs in $\posbody{\R}$\footnote{
In literature, ${\cal G}(\p)$ is also referred as {\em positive dependencies graph}.}.
A {\em component} $\,C$ of $\p$ is then a set of predicates which are strongly 
connected in ${\cal G}(\p)$.

The component graph of $\p$, denoted ${\cal G^C}(\p)$, is a labelled
directed graph having $(i)$ a node for each component of ${\cal G}(\p)$,
$(ii)$ an edge $C' \rightarrow^{\tt +} C$ if
there is a rule $\R \in \p$ such that a predicate ${\tt p} \in
C$ occurs in $\HR$ and a predicate ${\tt q} \in C'$
occurs in $\posbody{\R}$,
and $(iii)$ an edge $C' \rightarrow^{\tt -} C$ if (a) $C' \rightarrow^{\tt +} C$
is not an edge of ${\cal G^C}(\p)$, and (b)
there is a rule $\R \in \p$ such that a predicate ${\tt p} \in
C$ occurs in $\HR$ and a predicate ${\tt q} \in C'$
occurs in $\negbody{\R}$.
A path in a component graph ${\cal G^C}(\p)$ is {\em weak} if at least one of 
its edges is labelled with \dquo{$\tt -$}, otherwise it is {\em strong}.

A component ordering $\gamma = \langle C_1, \dots, C_n \rangle$ is a total 
ordering of all the components of $\p$ such that, for any $C_i$, $C_j$ with $i < j$, 
both $(a)$ there is no strong path from $C_j$ to $C_i$ in ${\cal G^C}(\p)$,
and $(b)$ if there is a weak path from $C_j$ to $C_i$, then there must be
a weak path also from $C_i$ to $C_j$.
A {\em module} $P(C_i)$ of a program $\p$ is the set of rules
defining predicates in $C_i$, excluding those
that define also some other predicate belonging to a lower
component in $\gamma$, that is, a component $C_j$ with $j < i$.

Given a rule $\R$ and a set $A$
of ground atoms, an instance $\R_g$ of $\R$
is an {\em $A$-restricted} instance of $\R$ if $\posbody{\R_g} \subseteq A$. 
The set of all $A$-restricted instances of all the rules of a program $\p$ is denoted
by $Inst_\p(A)$.
Note that, for any $A \subseteq \BP$, $Inst_\p(A) \subseteq
\GP$. Intuitively, this identifies those ground instances that may be
{\em supported} by a given set $A$.
\nop{Some further simplification to $\GP$ can be properly performed by
exploiting a modular evaluation of the program that relies on a
component ordering.}

Let $\p$ be a program, $C_i$ a component in a component ordering 
$\langle C_1,\ \ldots,\ C_n \rangle$, $T$ a set of ground rules to be 
simplified w.r.t.\ another set $R$ of ground rules.
Then the {\em simplification} $Simpl(T,R)$ of $T$ w.r.t.\ $R$ is obtained
from $T$ by: $(a)$ {\em deleting} each rule $\R_g$ such that 
$\head{\R_g} \cup \negbody{\R_g}$ contains some atom ${\tt p(\t)} \in \Facts(R)$; 
$(b)$ {\em eliminating}
from each remaining rule $\R_g$ the atoms in $\posbody{\R_g} \cap \Facts(R)$, 
and each atom ${\tt p(\t)} \in \negbody{\R_g}$ 
such that ${\tt p} \in C_j$, with $j < i$, and there is no rule in $R$ with
$\tt p(\t)$ in its head.
Assuming that $R$ contains all ground instances obtained from
the modules preceding $C_i$, $Simpl(T,R)$ deletes from $T$
the rules whose head is certainly already true w.r.t.\ $R$
or whose body is certainly false w.r.t.\ $R$, and
simplifies the remaining rules by removing from the bodies all
literals true w.r.t.\ $R$. We define now the operator $\Phi$,
combining $Inst$ and $Simpl$.


Let $\p$ be a program, $C_i$ a component in
a component ordering $\langle C_1,\ \ldots,\ C_n \rangle$, 
$R$ and $S$ two sets of ground rules.
Then $\Phi_{P(C_i),R}(S)= Simpl(Inst_{P(C_i)}(A), R)$,
where $A$ is the set of atoms belonging to the head of some rule in 
$R \cup S$.
The operator $\Phi$ always admit a least fixpoint 
$\Phi_{P(C_i),R}^\infty(\emptyset)$.
We can then define the {\em intelligent instantiation} $\p^\gamma$ of a program $\p$ 
for a component ordering $\gamma = \langle C_1,\ \ldots,\ C_n \rangle$ 
as the last element $\intelligent_n^\gamma$ of the sequence $\intelligent_0^\gamma = \EDB(\p)$,
$\intelligent_{i}^\gamma = \intelligent_{i-1}^\gamma \cup \Phi^\infty_{P(C_{i}),\intelligent_{i-1}^\gamma}(\emptyset)$. 
$\p$ is {\em finitely ground} (\fg)
if $\p^\gamma$ is finite for every component ordering~$\gamma$ for $\p$.
The main result for this class of programs is that reasoning is effectively computable.

\begin{theorem}\label{theo:fg-reasoningDecidable}
Cautious and brave
reasoning over \fg programs are decidable.
\end{theorem}

\subsection{Finitely Recursive Queries}\label{sec:magic_fr_queries}

We next provide the definition of finitely recursive queries \cite{cali-etal-2009-lpnmr}
and programs \cite{base-etal-2009-tplp}.

Let $\p$ be a program and $\q$ a query.
The relevant atoms for $\q$ are:
$(a)$ $\q$ itself, and $(b)$ each atom in $\atoms{\R_g}$, where $\R_g \in \GP$ is such that
some atom in $\head{\R_g}$ is relevant for $\q$.
Then $(i)$ $\q$ is {\em finitely recursive} on $\p$ if
only a finite number of ground atoms is relevant for $\q$,
and $(ii)$ $\p$ is {\em finitely recursive} if every query is finitely recursive on $\p$.

\begin{example}\label{ex:finitely_recursive}
Consider the query $\tt greaterThan(s(s(0)),0)?$ for the following program:
\begin{dlvcode}
\R_1:\quad \tt lessThan(X,s(X)). \\
\R_2:\quad \tt lessThan(X,s(Y)) \derives lessThan(X,Y). \\
\R_3:\quad \tt greaterThan(s(X),Y) \derives \naf~lessThan(X,Y).
\end{dlvcode}%
The program cautiously and bravely entails the query.
The query is clearly finitely recursive; also the program is finitely recursive.
$\mathproofbox$
\end{example}

\section{Magic-Set Techniques}\label{sec:magic}

The Magic Set method is a strategy for simulating the top-down
evaluation of a query by modifying the original program by means of
additional rules, which narrow the computation to what is relevant for
answering the query.  In this section we first recall the magic set
technique for disjunctive programs with stratified negation without
function symbols, as presented in
\cite{alvi-etal-2009-TR}, we then lift the technique to
\ASPFNFR{} queries, and
formally prove its correctness.

\subsection{Magic Sets for Function-Free Programs}
\label{sec:msFuncFree}

The method of \cite{alvi-etal-2009-TR}\footnote{For
a detailed description of the standard technique 
we refer to \cite{ullm-89}.} is structured in three main phases.

\noindent
\textbf{(1) Adornment.} The key idea is to materialize the binding information for IDB predicates that would be
propagated during a top-down computation,
like for instance the one adopted by Prolog. According to this kind of evaluation, 
all the rules $\R$ such that ${\tt g(\t')} \in \HR$ (where ${\tt g(\t')}\vartheta = \Q$ for some
substitution $\vartheta$) 
are considered in a first step. Then the atoms in $\atoms{\R\vartheta}$ 
different from $\Q$ are considered as new queries and the procedure is iterated.

Note that during this process the information about \emph{bound}
(i.e.\ non-variable) arguments in the query is ``passed'' to the other
atoms in the rule. Moreover, it is assumed that the rule is processed in
a certain sequence, and processing an atom may bind some of its
arguments for subsequently considered atoms, thus ``generating'' and
``passing'' bindings.  Therefore, whenever an atom is processed, each
of its arguments is considered to be either \emph{bound} or
\emph{free}.

The specific propagation strategy adopted in a top-down evaluation scheme is called {\em sideways information
passing strategy} (SIPS), which is just a way of formalizing a partial ordering over the atoms of each rule
together with the specification of how the bindings originated and propagate
\cite{beer-rama-91,grec-2003}.
Thus, in this phase, adornments are first created for the query predicate.
Then each adorned predicate is used to propagate its information to the other atoms of the rules defining it
according to a SIPS, thereby simulating a top-down evaluation. 
While adorning rules, novel binding information in the form of yet unseen adorned predicates may be generated, which should be used
for adorning other rules.

\noindent
\textbf{(2) Generation.} The adorned rules are then used to generate
{\em magic rules} defining {\em magic predicates}, which represent the atoms relevant for answering the input query.
Thus, the bodies of magic rules contain the atoms required for binding
the arguments of some atom, following the adopted SIPS.

\noindent
\textbf{(3) Modification.} Subsequently, magic atoms are added to the bodies of the adorned rules in order to
 limit the range of the head variables, thus avoiding the inference of facts which are irrelevant for the query. The resulting rules are called {\em modified rules}.

The complete rewritten program consists of the magic and modified rules
(together with the original EDB).  Given a function-free program $\P$, a 
query $\Q$, and the rewritten program $\P'$, $\P$
and $\P'$ are equivalent w.r.t.\ $\Q$, i.e., $\P\bqequiv{\Q} {\P'}$ and $\P\cqequiv{\Q} {\P'}$ hold
\cite{alvi-etal-2009-TR}.

\subsection{A Rewriting Algorithm for \ASPFNFR{} Programs}
\label{sec:finirec-magic}

Our rewriting algorithm exploits the peculiarities of 
\ASPFNFR{} queries, and guarantees that the rewritten program is query
equivalent, that it has a particular structure and that it is bottom-up
computable.
In particular, for a finitely recursive query $\Q$ over an \ASPFN{} program
$\p$, the Magic-Set technique can be simplified due to the following observations:

\begin{itemize}
 \item 
\vspace{-0.2cm}
  For each (sub)query $\tt g(\t)$ and each rule $\R$ having an atom ${\tt g(\t')} \in \HR$,
  all the variables appearing in $\R$ appear also in $\tt g(\t')$.
  Indeed, if this is not the case, then an infinite number of ground atoms 
  would be relevant for $\q$ (the query would not be finitely recursive).%
\footnote{We assume the general case where there is some functor with arity
greater than 0.}
  Therefore, each adorned predicate generated in the \textbf{Adornment} phase
  has all arguments bound.
  
 \item
  Since all variables of a processed rule are bound by the (sub)query,
  the body of a magic rule produced in the \textbf{Generation} phase
  consists only of the magic version of the
  (sub)query (by properly limiting the adopted SIPS).
\vspace{-0.2cm}
\end{itemize}

\noindent
We assume the original program 
has no predicate symbol that 
begins with the string ``$\tt magic\_$''. In the following we will then use
$\tt magic\_p$ for denoting the magic predicate associated with the
predicate $\tt p$. So the magic atom associated with $\tt p(\t)$ 
will be $\tt magic\_p(\t)$, in which, by previous considerations, each argument is bound.

\begin{figure}[t]
 \centering
 \figrule
 \mbox{\hspace{2mm}\parbox{0.88\textwidth}{\scriptsize
  \begin{description}
  \item[Input:] A program $\P$, and a query $\Q = \tt g(\t)?$
  \item[Output:] The optimized program $\DMS(\Q,\P)$.
  \item[var]  $S$, $D$: \textbf{set} of predicates;\ \ \modifiedRules$_{\Q,\P}$, \magicRules$_{\Q,\P}$: \textbf{set} of rules;
  \item[begin] \
  \item[]\emph{\phantom{0}1.}\ \ $D$ := $\emptyset$;\ \ \modifiedRules$_{\Q,\P}$ := $\emptyset$;\ \ \magicRules$_{\Q,\P}$ := $\{{\tt magic\_g(\t).}\}$;\ \ $S$ := $\{{\tt g}\}$;
  \item[]\emph{\phantom{0}2.}\ \ \textbf{while} $S\neq \emptyset$ \textbf{do}
  \item[]\emph{\phantom{0}3.}\ \ \hs take an element $\tt p$ from $S$; $\quad$ remove $\tt p$ from $S$; $\quad$ add $\tt p$ to $D$;
  \item[]\emph{\phantom{0}4.}\ \ \hs \textbf{for each} rule $\R \in \P$ and \textbf{for each} atom $\tt p(\t)$ $\in \HR$ \textbf{do}
  \item[]\emph{\phantom{0}5.}\, \ \hs \hs $\R'$ := $\R$;
  \item[]\emph{\phantom{0}6.}\, \ \hs \hs \textbf{for each} atom ${\tt q(\s)} \in \HR$ \textbf{do} \hs add ${\tt magic\_q(\s)}$ to $\body{\R'}$; \hs \textbf{end for}
  \item[]\emph{\phantom{0}7.}\, \ \hs \hs add $\R'$ to \modifiedRules$_{\Q,\P}$;
  \item[]\emph{\phantom{0}8.}\ \ \hs \hs \textbf{for each} atom ${\tt q(\s)} \in \atoms{\R} \setminus \{{\tt p(\t)}\}$ such that $\tt q$ is an IDB predicate \textbf{do}
  \item[]\emph{\phantom{0}9.}\ \ \hs \hs \hs add ${\tt magic\_q(\s) \derives magic\_p(\t).}$ to \magicRules$_{\Q,\P}$; \hs add ${\tt q}$ to $S$ \textbf{if} ${\tt q} \not\in D$;
  \item[]\emph{10.}\ \ \hs \hs \textbf{end for}
  \item[]\emph{11.}\ \ \hs \textbf{end for}
  \item[]\emph{12.}\ \ \textbf{end while}
  \item[]\emph{13.}  $\DMS(\Q,\P)$ := \magicRules$_{\Q,\P}$ \ $\cup$ \modifiedRules$_{\Q,\P}$ $\cup$ \EDB\!\!(\p);
  \item[]\emph{14.}  \textbf{return} $\DMS(\Q,\P)$;
  \item[end.] \
  \end{description}
 }}
 \caption{Magic Set algorithm ($\DMS$) for \ASPFNFR{} queries.}\label{fig:DMS}
 \figrule
\end{figure}

The algorithm $\DMS$ implementing the Magic-Set technique for \ASPFNFR{}
queries is reported in Figure~\ref{fig:DMS}.
Given a program $\p$ and a query $\Q$, the algorithm outputs a rewritten and
optimized program $\DMS(\Q,\p)$, consisting of a set of
\emph{modified} and \emph{magic} rules, stored by means of the sets 
\modifiedRules$_{\Q,\P}$ and \magicRules$_{\Q,\P}$, respectively
(together with the original EDB). 
The algorithm exploits a set $S$ for storing all the predicates to
be processed, and a set $D$ for storing the predicates already done. 

The computation starts by initializing $D$ and \modifiedRules$_{\Q,\P}$ to the empty set
(step \emph{1}). Then the
magic seed $\tt magic\_g(\t).$ (a fact) is stored in \magicRules$_{\Q,\P}$ and
the predicate $\tt g$ is inserted in the set $S$ (step \emph{1}). 
The core of the algorithm (steps \emph{2--12}) is repeated until the set $S$ is empty, i.e., until there is no
further predicate to be propagated. In particular, a predicate $\tt p$ is moved from
$S$ to $D$ (step \emph{3}), and each rule $\R \in \P$ 
having an atom $\tt p(\t)$ in the head is considered (note that one rule $\R$ is
processed as often as $\tt p$ occurs in its head; steps \emph{4--11}).
A modified rule $\R'$ is subsequently obtained from $\R$ by adding 
an atom $\tt magic\_q(\s)$ (for each atom $\tt q(\s)$ in the head of $\R$)
to its body (steps \emph{5--7}).
In addition, for each atom $\tt q(\s)$ in $\atoms{\R} \setminus \{{\tt p(\t)}\}$
such that $\tt q$ is an IDB predicate (steps \emph{8--10}), a magic rule 
$\tt magic\_q(\s) \derives magic\_p(\t).$ is generated (step \emph{9}), and the predicate
$\tt q$ is added to the set $S$ if not already processed (i.e., if ${\tt q} \not\in D$; step \emph{9}).
Note that the magic rule $\tt magic\_q(\s) \derives magic\_p(\t).$ is added
also if $\tt q(\s)$ occurs in the head or in the negative body,
since bindings are propagated in a uniform way to all IDB atoms.

\begin{example}
The result of the application of the $\DMS$ algorithm to the program and query 
in Example~\ref{ex:finitely_recursive} is:
\begin{dlvcode}
\R_1'\,:\quad \tt lessThan(X,s(X)) \derives magic\_lessThan(X,s(X)). \\
\R_2'\,:\quad \tt lessThan(X,s(Y)) \derives magic\_lessThan(X,s(Y)),\ lessThan(X,Y). \\
\R_3'\,:\quad \tt greaterThan(s(X),Y) \derives magic\_greaterThan(s(X),Y),\ \naf~lessThan(X,Y). \\
\R_2^\magica\,:\quad \tt magic\_lessThan(X,Y) \derives magic\_lessThan(X,s(Y)). \\
\R_3^\magica\,:\quad \tt magic\_lessThan(X,Y) \derives magic\_greaterThan(s(X),Y).\\
\R_{\Q}:\quad \tt magic\_greaterThan(s(s(0)),0).\mathproofbox
\end{dlvcode}%
\end{example}

\subsection{Query Equivalence Result}\label{sec:teoria}

We conclude the presentation of the $\DMS$ algorithm by formally proving its correctness. This section essentially follows \cite{alvi-etal-2009-TR}, to which we refer for the details, while here we highlight the necessary considerations for generalizing the results of \cite{alvi-etal-2009-TR} to \ASPFNFR{} queries, exploiting the considerations described in Section~\ref{sec:finirec-magic}. Throughout this section, we
use the well established notion of unfounded set for disjunctive programs with negation
defined in \cite{leon-etal-97b}. Since we deal with total interpretations, represented as the set of atoms interpreted as true, the
definition of unfounded set can be restated as follows.

\begin{definition}[Unfounded sets]
\label{def:unfoundedset} Let $I$ be an interpretation for a program $\p$, and $X \subseteq \BP$
be a set of ground atoms. Then $X$ is an \emph{unfounded set} for $\p$ w.r.t.\ $I$ if and only if for each ground rule
$\R_g \in \ground{\p}$ with $X \cap \head{\R_g} \neq \emptyset$, either $(1.a)$ $\posbody{\R_g} \not\subseteq I$, 
or $(1.b)$ $\negbody{\R_g} \cap I \neq \emptyset$, or
$(2)$ $\posbody{\R_g} \cap X \neq \emptyset$, or $(3)$ $\head{\R_g} \cap (I \setminus X) \neq \emptyset$.
\end{definition}

Intuitively, conditions $(1.a)$, $(1.b)$ and $(3)$ check if the rule is satisfied by $I$ regardless of the atoms in $X$,
while condition $(2)$ assures that the rule can be satisfied by taking the atoms in $X$ as false.
Therefore, the next theorem immediately follows from the characterization of
unfounded sets in~\cite{leon-etal-97b}.

\begin{theorem}\label{theo:unfounded}
Let $I$ be an interpretation for a program $\p$. Then, for any stable 
model $M \supseteq I$ of $\p$, and for each unfounded set $X$ of $\p$ w.r.t.\ $I$, 
$M \cap X = \emptyset$ holds. 
\end{theorem}

We now prove the correctness of the $\DMS$ strategy by showing that it is
\emph{sound} and \emph{complete}.
In both parts of the proof, we exploit the following set of atoms.

\begin{definition}[Killed atoms]
\label{def:killed} Given a model $M$ for $\dmsqp$, and a model $N \subseteq M$ of $\ground{\dmsqp}^{M}$, 
the set $\killedmn$ of the \emph{killed atoms}
w.r.t.\ $M$ and $N$ is defined as: 
$$
\{\,{\tt p(\t)} \in \BP \setminus N \ | \ \mbox{ either }\, {\tt p}\, 
    \mbox{ is an EDB predicate, or } {\tt magic\_p(\t)} \in N\,\}.
$$
\end{definition}

Thus, killed atoms are either 
false instances of some EDB predicate, or false atoms which are relevant 
for $\Q$ (since a magic atom exists in $N$).
Therefore, we expect that these atoms are also false in any stable model for $\p$ 
containing $M \cap \BP$. 
%

\begin{proposition}
\label{prop:killed_unfounded} Let $M$ be a model for $\dmsqp$, 
and $N \subseteq M$ a model of $\ground{\dmsqp}^{M}$. Then $\killedmn$ is an unfounded set
for $\p$ w.r.t.\ $M \cap \BP$.
\end{proposition}

We can now prove the soundness of the algorithm.

\nop{
\begin{lemma}
\label{lem:one_minimal_model} 
Let $\Q$ be a query for a stratified program $\p$. Then,
for each stable model $M'$ of $\dmsqp$, there is a stable model $M$ of $\p$ such
that $M \supseteq M' \cap \BP$.
\end{lemma}
\begin{proof}
Let $M'$ be a stable model of $\dmsqp$.
Consider the program $\p \cup \Mpp$ and note that $\killedmpmp$ is an unfounded set for it.
Let $M$ be a stable model for $\p \cup \Mpp$ and suppose, by contradiction,
there is a model $N \subset M$ of $\ground{\p}^M$.
We want to show that $N' = M' \setminus (M \setminus N)$ is a model of 
$\ground{\dmsqp}^{M'}$.
To this end, consider a modified rule $\R_g' \in \ground{\dmsqp}$ such that
$\posbody{\R_g'} \subseteq N'$ and $\negbody{\R_g'} \cap M' = \emptyset$:
\begin{dlvcode}
\R_g': \ \tt p_1(\t_1)\,\Or\,\cdots\,\Or\,p_n(\t_n) \derives 
     magic\_p_1(\t_1),\,\ldots,\,magic\_p_n(\t_n),\\
\phantom{\R_g': \ \tt p_1(\t_1)\,\Or\,\cdots\,\Or\,p_n(\t_n) \derives }
     \tt q_1(\s_1),\,\ldots,\,q_j(\s_j),\,\naf~q_{j+1}(\s_{j+1}),\,\ldots,\,\naf~q_m(\s_m).
\end{dlvcode}%
Thus, by Lemma~\ref{lem:mappingGroundNonground},
\begin{dlvcode}
\R_g: \ \tt p_1(\t_1)\,\Or\,\cdots\,\Or\,p_n(\t_n) \derives 
    q_1(\s_1),\,\ldots,\,q_j(\s_j),\,\naf~q_{j+1}(\s_{j+1}),\,\ldots,\,\naf~q_m(\s_m).
\end{dlvcode}%
belongs to $\GP$.
Note that each atom in $\head{\R_g'} \cup \negbody{\R_g'}$ which is false
w.r.t.\ $M'$ belongs to $\killedmpmp$. Thus, we have $\negbody{\R_g} \cap M = \emptyset$
and $\posbody{\R_g} \subseteq N$.
Moreover, $\head{\R_g} \cap N = \head{\R_g'} \cap N' \neq \emptyset$.
\end{proof}
}

\begin{lemma}
\label{lem:extending_minimal_models} Let $\Q$ be an \ASPFNFR{} query over $\p$. Then, for each stable
model $M'$ of $\dmsqp$, there is a stable model $M$ of $\p$ such that 
$\Q \in M$ if and only if $\Q \in M'$.
\end{lemma}
\begin{proof}
We can show that there is $M \in \SM(\p)$ 
such that $M\supseteq M' \cap \BP$.
Since $\Q$ belongs either to $M'$ or to $\killedmpmp$, 
the claim follows by Proposition~\ref{prop:killed_unfounded}.
\end{proof}

For proving the completeness of the algorithm we provide a construction for passing from an interpretation for $\p$ to one for $\dmsqp$.

\begin{definition}[Magic variant]
\label{def:magic_variant} Let $I$ be an interpretation for an \ASPFNFR{} query $\Q$ over $\p$. We define an interpretation $\variantqpi$ for
$\dmsqp$, called the magic variant of $I$ w.r.t.\ $\Q$ and $\p$, as follows:
$$
\variantqpi = \EDB(\p) \cup M^\magica \cup \{ {\tt p(\t)} \in I \ \mid {\tt magic\_p(\t)} \in M^\magica \},
$$
\noindent
where $M^\magica$ is the unique stable model of \magicRules$_{\Q,\P}$.
\end{definition}

In this definition, we exploit the fact that \magicRules$_{\Q,\P}$ has a unique and finite stable model for \ASPFNFR{} queries (see Lemma~\ref{lem:magicUnique} for a detailed proof).
By definition, for a magic variant $\variantqpi$ of an interpretation $I$ for $\p$, $\variantqpi \cap \BP \subseteq
I$ holds. More interesting, the magic variant of a stable model for $\p$ is in turn a stable model for $\dmsqp$
preserving truth/falsity of $\Q$.
The following formalizes the intuition above.

\begin{lemma}\label{lem:variant_stable_model} 
If $M$ is a stable model of an \ASPFN{} program $\p$ with a finitely recursive query $\Q$, then $M' = \variantqpm$ is a stable model of $\dmsqp$
and $\Q \in M'$ if and only if $\Q \in M$.
\end{lemma}
\begin{proof}
Consider a modified rule $\R_g' \in \ground{\dmsqp}$ having
$\posbody{\R_g'} \subseteq M'$ and $\negbody{\R_g'} \cap M' = \emptyset$:
\begin{dlvcode}
\R_g': \ \tt p_1(\t_1)\,\Or\,\cdots\,\Or\,p_n(\t_n) \derives 
     magic\_p_1(\t_1),\,\ldots,\,magic\_p_n(\t_n),\\
\phantom{\R_g': \ \tt p_1(\t_1)\,\Or\,\cdots\,\Or\,p_n(\t_n) \derives }
     \tt q_1(\s_1),\,\ldots,\,q_j(\s_j),\,\naf~q_{j+1}(\s_{j+1}),\,\ldots,\,\naf~q_m(\s_m).
\end{dlvcode}%
We can show that
\begin{dlvcode}
\R_g: \ \tt p_1(\t_1)\,\Or\,\cdots\,\Or\,p_n(\t_n) \derives 
    q_1(\s_1),\,\ldots,\,q_j(\s_j),\,\naf~q_{j+1}(\s_{j+1}),\,\ldots,\,\naf~q_m(\s_m).
\end{dlvcode}%
belongs to $\ground{\p}$. 
\nop{
\begin{proof}
By definition, $\R_g \in \ground{\dmsqp}$ if and only if there is $\R \in \p$ 
such that $\R_g = \R\vartheta$ for some substitution $\vartheta$.
Since ${\tt magic\_p_i(\t_i)} \in \ensuremath{B_{\dmsqp}}$, $\R \in \p$ and
$\R_g = \R\vartheta$
if and only if a modified rule $\R'$ such that $\R_g' = \R'\vartheta$ has been
produced.
\end{proof}
}
Since $\posbody{\R_g'} \subseteq M'$ and 
$\negbody{\R_g'} \cap M' = \emptyset$, we have 
$\posbody{\R_g} \subseteq M$, $\negbody{\R_g} \cap M = \emptyset$,
and $\head{\R_g'} \cap M' = \head{\R_g} \cap M$. 
Thus, $\head{\R_g'} \cap M' = \head{\R_g} \cap M \neq \emptyset$ because
$M$ is a model of $\p$.
Moreover, if there is a model $N' \subset M'$ of $\ground{\dmsqp}^{M'}$,
then $M \setminus (M' \setminus N')$ is a model for $\ground{\p}^M$,
contradicting the assumption that $M$ is a stable model of $\p$.

Thus, $M' = \variantqpm$ is a stable
model of $\dmsqp$.
Since $\Q$ belongs either to $M'$ or to $\killedmpmp$, 
the claim follows by Proposition~\ref{prop:killed_unfounded}.
\end{proof}

From the above lemma, together with Lemma~\ref{lem:extending_minimal_models}, the
correctness of the Magic Set method with respect to query answering directly follows.

\begin{theorem}\label{thm:equivalence}
\label{theo:dms_equivalence}
If  $\Q$ is an \ASPFNFR{} query over $\p$, then both $\dmsqp \bqequiv{\Q}
\p$ and $\dmsqp \cqequiv{\Q} \p$ hold.
\end{theorem}

\section{Decidability Result}\label{sec:decidability}

In this section, we prove that \ASPFNFR{} queries are decidable.
To this end, we link finitely recursive queries to finitely ground programs.
More specifically, we show that the Magic-Set rewriting of a finitely recursive
query is a finitely ground program, for which querying is known to be decidable.

We first show some properties of the rewritten program due to the particular
restrictions applied to the adopted SIPS.

\begin{lemma}\label{lem:magicStratified}
If $\Q$ is an \ASPFNFR{} query over $\p$, 
then $\DMS(\Q,\p)$ is stratified.
\end{lemma}
\begin{proof}
Each cycle of
dependencies in $\dmsqp$ involving predicates of $\p$ is also present in
$\p$. Indeed, each magic rule has exactly one magic atom in the head and one
in the body, and each modified rule is obtained by adding 
only magic atoms to the body of a rule belonging to $\p$. 
Since $\p$ is stratified by assumption, such cycles have 
no negative dependencies.
Any new cycle stems only from magic rules, which are positive.
\end{proof}

Now consider the program consisting of the magic rules produced for 
a finitely recursive query. 
We can show that this program has a unique and finite stable model,
that we will denote $M^\magica$.

\begin{lemma}\label{lem:magicUnique}
Let $\Q$ be an \ASPFNFR{} query over $\p$.
Then the program \magicRules$_{\Q,\p}$ has a unique and finite stable model $M^\magica$.
\end{lemma}
\begin{proof}
Since \magicRules$_{\Q,\p}$ is positive and normal, $M^\magica$ is unique.
If we show that $M^\magica$ contains all and only the relevant atoms for $\Q$,
then we are done because $\Q$ is finitely recursive on $\p$.
To this end, note that the only fact in \magicRules$_{\Q,\p}$ is the query seed
$\tt magic\_g(\t).$, and each magic rule $\tt magic\_q(\s)\vartheta \derives magic\_p(\t)\vartheta.$
in $\ground{\dmsqp}$ ($\vartheta$ a substitution) is such that
${\tt q(\s)}\vartheta$ is relevant for ${\tt p(\t)}\vartheta$.
Indeed, $\tt magic\_q(\s) \derives magic\_p(\t).$ has been produced during the 
{\em Generation} phase involving a rule $\R \in \p$ with ${\tt p(\t)} \in \HR$ 
and ${\tt q(\s)} \in \atoms{\R} \setminus \{{\tt p(\t)}\}$;
since each variable in $\R$ appears also in $\tt p(\t)$,
$\R\vartheta \in \ground{\p}$ is such that ${\tt p(\t)}\vartheta \in \head{\R\vartheta}$ and
${\tt q(\s)}\vartheta \in \atoms{\R\vartheta}$, i.e., ${\tt q(\s)}\vartheta$
is relevant for ${\tt p(\t)}\vartheta$.
\end{proof}

We can now link \ASPFNFR{} queries and finitely ground programs.

\begin{theorem}\label{theo:magicFinitelyGround}
Let $\Q$ be an \ASPFNFR{} query over $\p$.
Then $\DMS(\Q,\p)$ is finitely ground.
\end{theorem}
\begin{proof}
Let $\gamma = \langle C_1, \ldots, C_n \rangle$ be a component ordering for
$\dmsqp$.
Since each cycle of dependencies in $\dmsqp$ involving predicates of $\p$
is also present in $\p$, components with
non-magic predicates are disjoint from components with magic predicates.
For a component $C_i$ with magic predicates, 
$\dmsqp_i^\gamma$ is a subset of $M^\magica$,
which is finite by Lemma~\ref{lem:magicUnique}.

For a component $C_i$ with a non-magic predicate $\tt p_u$, we consider
a modified rule $\R' \in P(C_i)$ with an atom ${\tt p_u(\t_u)} \in \head{\R'}$:
\begin{dlvcode}
\R': \ \tt p_1(\t_1)\,\Or\,\cdots\,\Or\,p_n(\t_n) \derives 
     magic\_p_1(\t_1),\,\ldots,\,magic\_p_n(\t_n),\\
\phantom{\R': \ \tt p_1(\t_1)\,\Or\,\cdots\,\Or\,p_n(\t_n) \derives }
     \tt q_1(\s_1),\,\ldots,\,q_j(\s_j),\,\naf~q_{j+1}(\s_{j+1}),\,\ldots,\,\naf~q_m(\s_m).
\end{dlvcode}%
Thus, the component containing $\tt magic\_p_u$ precedes $C_i$ in $\gamma$.
Moreover, since $\Q$ is finitely recursive on $\p$, each variable appearing in
$\R'$ appears also in $\tt magic\_p_u(\t_u)$.
Therefore, $\dmsqp_i^\gamma$ is finite also in this case.
\end{proof}

We are now ready for proving the decidability of brave and cautious reasoning
for the class of finitely recursive queries on \ASPFN{} programs.

\begin{theorem}\label{theo:decidability}
Let $\Q$ be an \ASPFNFR{} query over $\p$.
Deciding whether $\p$ cautiously/bravely entails $\Q$ is computable.
\end{theorem}
\begin{proof}
From Theorem~\ref{thm:equivalence}, 
$\dmsqp \bqequiv{\Q}\p$ and $\dmsqp \cqequiv{\Q} \p$ hold.
Since $\dmsqp$ is finitely ground by Theorem~\ref{theo:magicFinitelyGround},
decidability follows from Thereom~\ref{theo:fg-reasoningDecidable}.
\end{proof}

\section{Expressiveness Result}\label{sec:expressiveness}

In this section, we show that the restrictions which guarantee the
decidability of \ASPFNFR{} queries do not limit their expressiveness.
Indeed, any computable function can be encoded by an \ASPFNFR{}
program (even without using disjunction and negation).  To this end,
we show how to encode a deterministic Turing Machine as a positive
program with functions and an input string by means of a query.  In
fact it is well-known that Horn clauses (under the classic first-order
semantics) can represent any computable function \cite{tarn-77}, so we
just have to adapt these results for \ASPFNFR{} programs and queries.

A Turing Machine ${\cal M}$ with semi-infinite tape 
is a 5-tuple $\tuple{\Sigma, {\cal S}, {\tt s_i}, {\tt s_f}, \delta}$,
where $\Sigma$ is an alphabet (i.e., a set of symbols),
$\cal S$ is a set of states,
${\tt s_i}, {\tt s_f} \in \cal S$ are two distinct states (representing the initial and 
final states of $\cal M$, respectively), and
$\delta : {\cal S} \times \Sigma \longrightarrow {\cal S} \times \Sigma \times \{\leftarrow, \rightarrow\}$ 
is a transition function.
Given an input string $x = \tt x_1 \cdots x_n$, the initial configuration of $\cal M$
is such that the current state is $\tt s_i$,
the tape contains $x$ followed by an infinite sequence of blank symbols $\blank$
(a special tape symbol occurring in $\Sigma$; we are assuming $x$ does not contain
any blank symbol), 
and the head is over the first symbol of the tape.
The other configurations assumed by $\cal M$ with input $x$ are then obtained by means 
of the transition function $\delta$:
If $\tt s$ and $\tt v$ are the current state and symbol, respectively,
and $\tt \delta(s,v) = (s',v',m)$, then $\cal M$ overwrites $\tt v$ with $\tt v'$,
moves its head according to ${\tt m} \in \{\leftarrow, \rightarrow\}$,
and changes its state to $\tt s'$.
$\cal M$ {\em accepts} $x$ if the final state $\tt s_f$ is reached
at some point of the computation.

A configuration of $\cal M$ can be encoded by an instance of $\tt conf(s,L,v,R)$,
where $\tt s$ is the current state, $\tt v$ the symbol under the head,
$\tt L$ the list of symbols on the left of the head in reverse order,
and $\tt R$ a finite list of symbols on the right of the head containing
at least all the non-blank symbols.
The query $\q_{{\cal M}(x)}$ representing the initial configuration
of $\cal M$ with input $x$ is
\begin{dlvcode}
\begin{array}{lcl}
\tt conf(s_i,[\ ],x_1,[x_2, \ldots, x_n])? && \tt \mbox{if } n > 0; \\
\tt conf(s_i,[\ ],\blank,[\ ])? && \tt \mbox{otherwise.}
\end{array}
\end{dlvcode}%
The program $\p_{\cal M}$ encoding $\cal M$ contains a rule
$\tt conf(s_f,L,V,R).$ representing the final state $\tt s_f$,
and a set of rules implementing the transition function $\delta$.
For each state ${\tt s} \in {\cal S} \setminus \{{\tt s_f}\}$ and for 
each symbol ${\tt v} \in \Sigma$, $\p_{\cal M}$ contains the following rules:
\begin{dlvcode}
\begin{array}{lcl}
\tt conf(s,[V|L],v,R) \derives conf(s',L,V,[v'|R]). && \tt \mbox{if } \delta(s,v) = (s',v',\leftarrow); \\
\tt conf(s,L,v,[V|R]) \derives conf(s',[v'|L],V,R). && \tt \mbox{if } \delta(s,v) = (s',v',\rightarrow); \\
\tt conf(s,L,v,[\ ]) \ \ \ \derives conf(s',[v'|L],\blank,[\ ]). && \tt \mbox{if } \delta(s,v) = (s',v',\rightarrow).
\end{array}
\end{dlvcode}%
Note that we do not explicitly represent the infinite sequence of blanks on the right of the tape; the last rule above effectively produces a blank whenever the head moves right of all explicitly represented symbols. The atoms therefore represent only the effectivley reached tape positions. We now show the correctness of $\p_{\cal M}$ and $\q_{{\cal M}(x)}$.

\begin{theorem}\label{theo:turingCorrectness}
The program $\p_{\cal M}$ bravely/cautiously
entails $\q_{{\cal M}(x)}$ if and only if $\cal M$ accepts $x$.
\end{theorem}
\begin{proof}[Proof Sketch]
$\p_{\cal M}$ bravely/cautiously entails $\q_{{\cal M}(x)}$
if and only if the unique stable model of $\p_{\cal M}$ contains
a sequence of atoms $\tt conf(\t_1), \ldots, conf(\t_m)$ such that
${\tt conf(\t_1)}$ is the query atom,
${\tt conf(\t_m)}$ is an instance of ${\tt conf(s_f,L,V,R)}$,
and there is a rule in $\ground{\p_{\cal M}}$ (implementing the transition
function of $\cal M$)
having ${\tt conf(\t_i)}$ in head and ${\tt conf(\t_{i+1})}$ in the body,
for each $\tt i = 1, \ldots, m-1$.
Since instances of $\tt conf(\t)$ represent configurations of $\cal M$,
the claim follows.
\end{proof}

We can now link computable sets (or functions) and finitely recursive queries.

\begin{theorem}\label{theo:turingFinRecQ}
Let $L$ be a computable set (or function).
Then, there is an \ASPFN{} program $\p$ such that, for each string $x$,
the query $\q_x$ is finitely recursive on $\p$,
and $\p$ cautiously/bravely entails $\q_x$ if and only if $x \in L$.
\end{theorem}
\begin{proof}
Let ${\cal M}$ be a Turing Machine computing $L$ and $\p_{\cal M}$ be the
program encoding $\cal M$. Program $\p_{\cal M}$ is clearly in
\ASPFN{} (actually, it is even negation-free).
By Theorem~\ref{theo:turingCorrectness}, it only remains to prove that 
$\q_{{\cal M}(x)}$ is finitely recursive on $\p_{\cal M}$.
By construction of $\p_{\cal M}$, for each ground atom 
$\tt conf(\t)$ in ${\cal B}_{\P_{\cal M}}$,
there is exactly one rule in $\ground{\p_{\cal M}}$ having $\tt conf(\t)$
in head.
This rule has at most one atom $\tt conf(\t')$ in its body, and
implements either the transition function or the final state of $\cal M$.
Thus, the atoms relevant for $\q_{{\cal M}(x)}$ are exactly the atoms
representing the configurations assumed by $\cal M$ with input $x$.
The claim then follows because $\cal M$ halts in a finite number of steps
by assumption.
\end{proof}

We note that when applying magic sets on the Turing machine encoding,
the magic predicates effectively encode all reachable configurations,
and a bottom-up evaluation of the magic program corresponds to a
simulation of the Turing machine. Hence only
encodings of Turing machine invocations that visit all (infinitely
many) tape cells are not finitely recursive. We also note
that recognizing whether an \ASPFN{} query or a program is finitely
recursive is RE-complete\footnote{That is, complete for the class of recursively enumerable decision problems.}.


\section{Related Work}\label{sec:relatedwork}

\newcommand{\ldl}{$\mathcal{LDL}$\xspace}
\newcommand{\fdnc}{$\mathbb{FDNC}$\xspace}
\newcommand\dependson{\geqslant}


The extension of ASP with functions has been the subject
of intensive research in the last years.
The main proposals can be classified in two groups:

1. {\em Syntactically restricted fragments}, such as
    $\omega${\em -restricted
    programs}~\cite{syrj-2001}, $\lambda${\em
    -restricted programs}~\cite{gebs-etal-2007-lpnmr},
    {\em finite-domain programs}
    \cite{cali-etal-2008-iclp}, {\em argument}-{\em
    restricted programs} \cite{lier-lifs-2009-iclp},
    \fdnc\ {\em programs} \cite{simk-eite-2007-lpar},
    {\em bidirectional programs}
    \cite{eite-simk-2009-ijcai}, and the proposal of
    \cite{lin-wang-2008-KR}; these approaches introduce
    syntactic constraints (which can be easily checked
    at small computational cost) or explicit domain
    restrictions, thus allowing computability of answer
    sets and/or decidability of querying;

2. {\em Semantically restricted fragments}, such as
    {\em finitely ground programs}~\cite{cali-etal-2008-iclp},
    {\em finitary programs}~\cite{bona-02-iclp,bona-04},
    {\em disjunctive finitely-recursive
    programs}~\cite{base-etal-2009-tplp} and {\em queries}~\cite{cali-etal-2009-lpnmr};
    with respect to syntactically restricted fragments, these 
    approaches aim at identifying broader classes of
    programs for which computational tasks such as
    querying are decidable. However, the membership of programs in these fragments is undecidable in general.

There have been a few other proposals that treat function symbols not
in the traditional LP sense, but as in classical
logic, where most prominently the unique names assumption does not
hold. We refer to \cite{caba-2008-iclp} for an overview.

Our work falls in the group 2. It is most closely related
to~\cite{bona-02-iclp},
\cite{base-etal-2009-tplp}, and especially \cite{cali-etal-2009-lpnmr},
which all focus on {\em querying} for {\em disjunctive} programs.

The work in~\cite{bona-02-iclp} studies how to extend
finitary programs \cite{bona-04} preserving decidability
for ground querying in the presence of disjunction. To this
end, an extra condition on disjunctive heads is added to the
original definition of finitary program of~\cite{bona-04}.
\nop{
Given a dependency relation which considers only
connections between head and body atoms (that is, $a
\dependson b$ iff there exists $r$ such that $a \in H(r)$
and $b \in B(r)$), a disjunctive program $P$ is finitary in
the sense of \cite{bona-02-iclp} if {\em (1)}~each ground
atom in $P$ depends on finitely many other atoms, {\em (2)}
the set $S$ of atoms appearing in odd-negated cycles is
finite and {\em (3)} the set $R$ of atoms $a$ for which
there is a rule $r \in P$ in which $a \in
max_{\dependson}(H(r))$ and there is an atom $b \in H(r)$
which is recursive with $a$ and $a$ positively depends on
$b$, is finite \footnote{Given erratum
\cite{bona-err-2008}, it turns out that both $S$ and $R$
must be known besides being finite.}.
}
Interestingly, the class of \ASPFNFR{} programs,
which features decidable reasoning (as proved in
Theorem~\ref{theo:decidability}), enlarges the stratified
subclass of disjunctive finitary programs
of~\cite{bona-02-iclp}. Indeed, while all stratified finitary
programs trivially belong to the class of \ASPFNFR{} programs,
the above mentioned extra condition on disjunctive heads is not guaranteed to be
fulfilled by \ASPFNFR{} programs
(even if negation is stratified or forbidden at all).
\nop{
as witnessed by the following program:
\begin{dlvcode2}
p(X) \Or q(X) \implied\ s(X).\qquad\qquad & q(X) \implied\ p(X). \\
p(f(X)) \implied\ q(X). & p(1). \\
p(X) \implied\ q(X). \\
\end{dlvcode2}
}
Instead, in \cite{base-etal-2009-tplp}, 
a redefinition (including disjunction) of
finitely recursive programs is considered, initially introduced
in~\cite{bona-04} as a super-class of finitary programs
allowing function symbols and negation. The authors show
a compactness property and semi-decidability results for cautious ground
querying, but no decidability results are given.


Our paper extends and generalizes the work \cite{cali-etal-2009-lpnmr},
in which the decidability of querying over finitely recursive {\em negation-free}
disjunctive programs is proved via a magic-set rewriting.
To achieve the extension, we had to generalize the magic set technique
used in \cite{cali-etal-2009-lpnmr} to deal also with stratified negation.
The feasibility of such a generalization was not obvious at all,
since the magic set rewriting of a stratified program can produce unstratified
negation~\cite{kemp-etal-95},
which can lead to undecidability 
in the presence of functions.
We have proved that, thanks to the structure of \ASPFNFR{} programs and the adopted SIPS,
the magic set rewriting preserves stratification.
The presence of negation also complicates the proof that the 
rewritten program is query-equivalent to the original one.
To demonstrate this result, we have exploited the characterization
of stable models via unfounded sets of \cite{leon-etal-97b},
and generalized the equivalence proof of \cite{alvi-etal-2009-TR}
to the case of programs with functions.

Finally, our studies on computable fragments of logic
programs with functions are loosely related to termination
studies of SLD-resolution for Prolog programs (see
e.g.~\cite{bruy-etal-2007-acm}).

\nop{
Some other papers about the magic-set
technique~\cite{banc-etal-1986,ullm-89,beer-rama-87}
are related to the present work as well, for which
different extensions and refinements have been proposed.
Among the more recent works, an adaptation for
soft-stratifiable programs~\cite{behr-2003-pods}, the
generalization to the disjunctive
case~\cite{cumb-etal-2004-iclp} and to Datalog with
(possibly unstratified) negation~\cite{fabe-etal-2007-jcss}
are worth remembering.
}

\section{Conclusion}\label{sec:conclusion}

In this work we have studied the language of \ASPFNFR{} queries and
programs. By adapting a magic set technique, any \ASPFNFR{} query can be
transformed into an equivalent query over a finitely ground
program, which is known to be decidable and for which an implemented
system is available. We have also shown that the \ASPFNFR{} language can
express any decidable function. In total, the proposed language and
techniques provide the means for a very expressive, yet decidable
and practically usable logic programming framework.

Concerning future work, we are working on adapting an existing
implementation of a magic set technique to handle \ASPFNFR{} queries
as described in this article, integrating it into
\dlv-Complex~\cite{dlvcomplex-web}, thus creating a useable \ASPFNFR{}
system. We also intend to explore practical application scenarios;
promising candidates are query answering over ontologies and in
particular the Semantic Web, reasoning about action and change, or
analysis of dynamic multi-agent systems.

\bibliography{bibtex}

\end{document}